\documentclass[12pt,a4paper]{article}
\pdfoutput=1
\usepackage{jheppub}
\usepackage[T1]{fontenc} 
\usepackage{epsfig} 
\usepackage{amscd}
\usepackage{verbatim}
\usepackage{latexsym}
\usepackage{amsfonts,amsthm,amsmath,mathtools}
\usepackage{amssymb,stmaryrd,textcomp,enumerate,bbm}
\usepackage{color}
\usepackage{xcolor}
\usepackage{booktabs}
\usepackage{float}
\usepackage{graphicx}
\usepackage{lscape}
\usepackage{slashed}
\notoc 

\title{\centering{
Notes on Integral Identities for  3d Supersymmetric Dualities
}}
\author[a]{Nezhla Aghaei}
\author[a,b]{\!\!,~Antonio Amariti}
\author[a]{\!\!,~Yuta Sekiguchi}
\affiliation[a]{Albert Einstein Center for Fundamental Physics, Institute for Theoretical Physics, University of Bern, Sidlerstrasse 5, Bern, ch-3012, Switzerland}
\affiliation[b]{INFN, Sezione di Milano, Via Celoria 16, I-20133 Milano, Italy}

\emailAdd{aghaei@itp.unibe.ch,antonio.amariti@mi.infn.it,yuta@itp.unibe.ch}

\newcommand{\tr}{{\rm tr}}

\newcommand{\nocontentsline}[3]{}
\newcommand{\tocless}[2]{\bgroup\let\addcontentsline=\nocontentsline#1{#2}\egroup}

\abstract{Four dimensional $\mathcal{N}=2$ Argyres-Douglas theories have been recently conjectured to 
be described by $\mathcal{N}=1$ Lagrangian theories. Such models, once reduced to 3d,
should be mirror dual to Lagrangian $\mathcal{N}=4$ theories. This has been numerically checked through the matching of the partition functions on the three sphere. In this article, we provide an analytic derivation for this result in the $A_{2n-1}$ case via hyperbolic hypergeometric integrals. We study the $D_4$ case as well, commenting on some open questions and possible resolutions.
In the second part of the paper we discuss other integral identities leading to
the matching of the partition functions in 3d dual pairs involving higher monopole superpotentials.
}

\begin{document}

\maketitle

\newpage
\tableofcontents
%
%
\section{Introduction}
%
%
It has been recently shown that 4d $\mathcal{N}=2$ Argyres-Douglas (AD) theories \cite{Argyres:1995jj}
can be obtained by an intricate RG flow structure 
\cite{Maruyoshi:2016tqk,Maruyoshi:2016aim,Agarwal:2016pjo,Benvenuti:2017lle,Benvenuti:2017kud,Benvenuti:2017bpg,Agarwal:2017roi}.
The analysis starts by considering a 4d $\mathcal{N}=2$ SCFT Lagrangian with a gauge group $G$ and hypermultiplets. 
Supersymmetry is broken to $\mathcal{N}=1$ by coupling the chiral multiplets 
with some singlets. A nilpotent vev for these singlets triggers an RG flow \cite{Maruyoshi:2016tqk,Maruyoshi:2016aim}.
In the IR a SCFT can be obtained by iterating a-maximization \cite{Intriligator:2003jj}, at the cost of introducing a set of accidental symmetries.
The properties of the SCFT under investigations are quite intriguing: it has been conjectured that there are
situations with rational central charge that in the IR enhance to AD theories\footnote{It has been shown that there are also 
cases with rational charges that do not enhance to $\mathcal{N}=2$ \cite{Evtikhiev:2017heo}.}.
This led to the conjecture that the $\mathcal{N}=1$ theory obtained in this way
 corresponds to the Lagrangian description
of AD theory.

An interesting consequence of the existence of such a Lagrangian formulation
is that one can reduce it to 3d and check if it 
reproduces correctly the expected
reduction of AD theories conjectured in \cite{Nanopoulos:2010bv}.
Substantial evidence
\footnote{We are grateful to Matthew Buican for precious comments on this issue.}
 for the conjectures of \cite{Nanopoulos:2010bv}
has been given in \cite{Buican:2015hsa}
by reducing the 4d superconformal index to the three sphere partition function.
In addition, in \cite{Buican:2015ina}
it was shown that expected RG flows following from the conjectured 3d quivers of \cite{Nanopoulos:2010bv}
were consistent with the form of the 4d index (see also \cite{Fredrickson:2017yka}). 
Finally, the latter result
has been extended to some "generalized" AD theories in \cite{Buican:2017uka}.

The idea has been recently pursued in \cite{Benvenuti:2017lle,Benvenuti:2017kud,Benvenuti:2017bpg} 
and it has been shown how to recover the results of
\cite{Nanopoulos:2010bv}  by reducing the 4d Lagrangian description of AD.
The main ideas that allow the authors to obtain the desired result are 
\emph{abelianization}, sequential confinement, and chiral ring stability \cite{Collins:2016icw}.
Many of these ideas are new and can potentially play a relevant role in the future analysis 
of 3d dualities.

\begin{figure}[H]
\begin{center}
  \includegraphics[width=12cm]{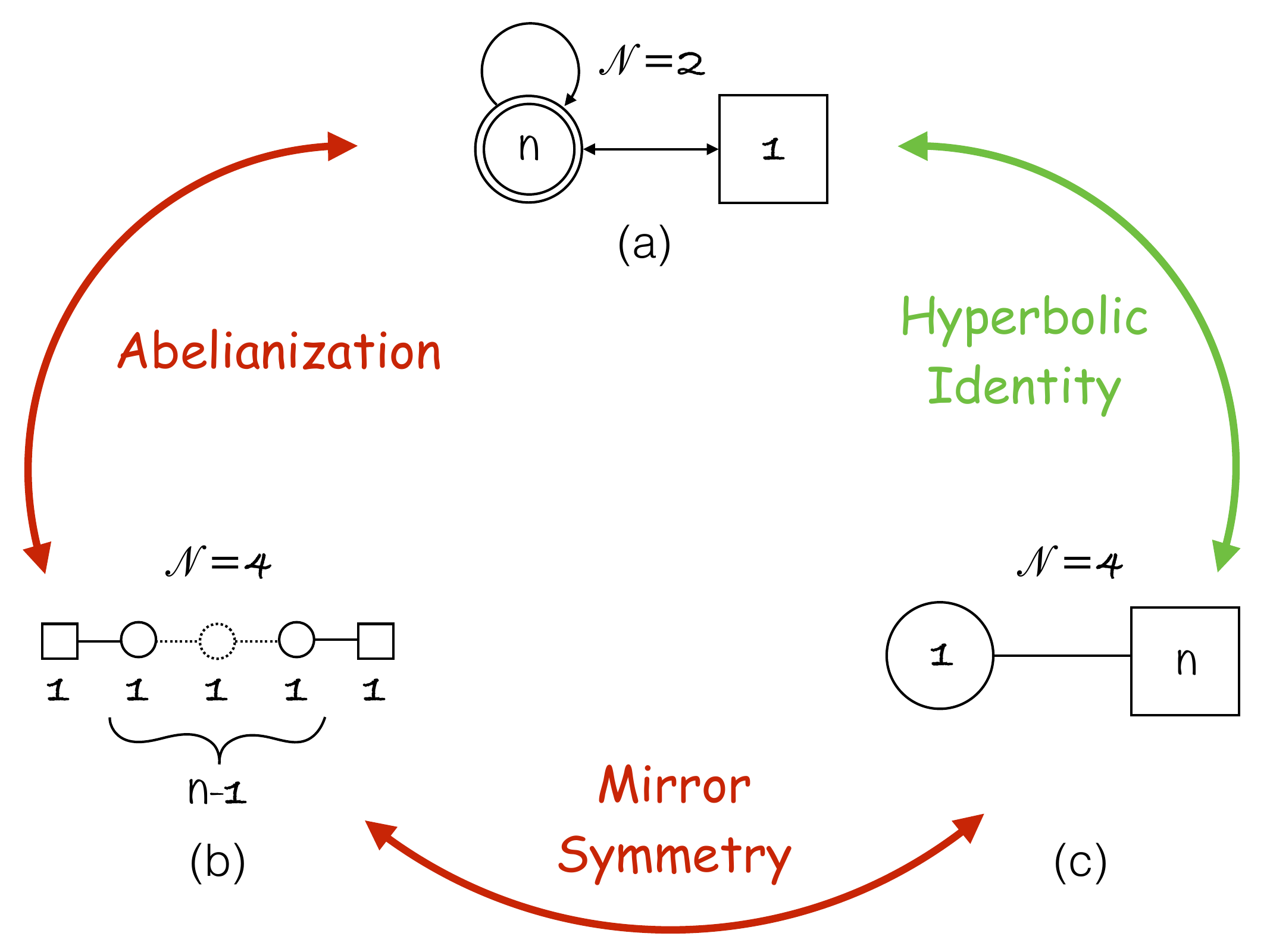}
  \caption{Relation between the relevant models describing the 3d $(A_1,A_{2n-1})$ AD theories
  reduced to 3d.
  We also specify the degree of supersymmetry of the quivers under consideration.
 The double circle denotes an $SU(n)$ gauge theory while single circles refer
 to unitary cases with abelian factors. Flavor symmetries are denoted by boxes.
    }
  \label{fig1}
\end{center}
\end{figure}

The relation between the Lagrangian description of the AD theories denoted as $(A_1,A_{2n-1})$
reduced to 3d and their 3d $\mathcal{N}=4$ mirror dual has been discussed in  \cite{Benvenuti:2017lle,Benvenuti:2017kud}.
It has been shown that these two theories can be mapped through an intermediate
step, where the natural $\mathcal{N}=4$ mirror quiver is
obtained through an \emph{abelianization} procedure.
In Figure \ref{fig1} we show this chain of dualities involving the three models.

The relation between models (a) and (b) in Figure \ref{fig1} has been numerically checked at the level of the partition function,
showing that for small $n$ (namely $n=2,3$) it is possible to prove that the two theories have the same
partition function providing a mapping between the R-charges and the real mass parameter \cite{Benvenuti:2017kud}.  
The relation is claimed to work for generic values of the real masses and charges, and after F-maximization,
the enhancement to $\mathcal{N}=4$ is expected. On the other hand, $\mathcal{N}=4$ mirror symmetry maps model (b) to (c), and the
equality between the $\mathcal{N}=4$ partition functions of the two models have been proven in \cite{Kapustin:2010xq}.

One starts from a possible UV completion, that reduces to the
model (a) in the IR after an RG flow and to the model (c) after mirror symmetry and a 
cascade of sequential confinements of the type discussed in \cite{Benini:2017dud}.
In \cite{Benvenuti:2017kud} it has been shown how to connect the two models (a) and (c) in Figure \ref{fig1} in an indirect way.
What is actually missing is a direct connection between models (a) and (c) of Figure \ref{fig1}.

In this paper we obtain this direct connection by exploiting some mathematical identities among hyperbolic
gamma functions and hyperbolic integrals. It turns out that the equivalence of the (squashed) three sphere 
partition functions of models (a) and (c) in Figure \ref{fig1} can be analytically proven for general $n$, hence
corroborating the results of \cite{Benvenuti:2017kud}.

The paper is organized as follows.
In section \ref{sec:hyp} we introduce the 
necessary main tools for our analysis, 
the representation
of the (squashed) three sphere partition function of 3d $\mathcal{N}=2$ theories in terms 
of hyperbolic gamma functions and hyperbolic hypergeometric integrals.
In section \ref{sec:a1an} we provide a quick review of the derivation of 4d $\mathcal{N}=1$ Lagrangian
conjectured to enhance to  $(A_1,A_{2n-1})$ in the IR, their reduction to 3d, the abelianization and the mirror
description. Then we derive the main result of this paper, the analytic matching of the three sphere
partition function between the reduced $(A_1,A_{2n-1})$ model and its mirror 3d $\mathcal{N}=4$ 
theory with $U(1)$ gauge group and $n$ hypermultiplets.
In section \ref{sec:D4} we speculate on the analogous result for the $(A_1,D_4)$ case. In this case abelianization
works in a different way and we have not been able to provide any analytical proof untill now. We explain the nature 
of the problem and propose another dual description that may play a useful role in the analysis.
In section \ref{sec:further} we discuss a further application of the identities among hyperbolic hypergeometric integrals.
This is related to some integral identities for the theories with symplectic gauge groups with
antisymmetric and fundamental matter . We restrict our analysis to the $Sp(2)$ case, when the antisymmetric matter disappears.
In the subsection \ref{subsec:multi} we show that if we start from the identities among the 3d hyperbolic integrals, it is possible to match the 
dual phases  which discussed in \cite{Benini:2017dud}, involving higher powers in the monopole superpotential.
In the subsection \ref{subsec:limaha} we show that considering the further real mass flow one can recover the limiting case of the usual Aharony duality
\cite{Aharony:1997gp} for the $U(2)$ theory with two flavors.
In section \ref{sec:conc} we conclude summarizing the main results and mention the open questions. 

\section{Hyperbolic integrals and partition function}
\label{sec:hyp}
In this section we review the mathematical formalism of hyperbolic hypergeometric integrals. Relation of these integrals 
with the partition function of 3d $\mathcal{N}=2$ supersymmetric
gauge theories computed from localization on the squashed three sphere $S_b^3$, where
$b$ represents the squashing parameter.
The partition function corresponds to a matrix integral over the real scalar $\sigma$ of the $\mathcal{N}=2$
vector multiplet, in the Cartan 
of the gauge group $G$ \cite{Jafferis:2010un,Hama:2010av,Hama:2011ea}.
It has been shown that the 1-loop contributions of the vector and of the matter fields
to the partition function can be formulated in terms of hyperbolic Gamma
functions $\Gamma_h(x)$ represented as follows
\begin{equation}
\label{eq:Gammahvbd}
 \Gamma_h(x) \equiv
\prod_{m,l=1}^{\infty}
\frac{(m+1)\,\omega_1+(l+1) \, \omega_2-x }{m \, \omega_1+l \, \omega_2+x}.
\end{equation}
Let us consider one example that will play a prominent role in our analysis,
a $U(n)$ gauge theory with $f$ pairs of fundamentals and anti--fundamental flavors and one adjoint.
The three sphere partition function of this model corresponds to the following matrix integral 
\begin{equation}
\label{Zgen} 
\frac{\Gamma_h(\tau)^{n}}{n !} \int \prod_{i=1}^{n} \frac{d \sigma_i}{\sqrt{-\omega_1 \omega_2}} \;
 e^{\frac{2i \pi \lambda \sigma_i}{\omega_1 \omega_2} }\;
  \!\!  \!\!  \!\!  \!\! 
 \prod_{1 \le i < j \le n } \!\! \frac{\Gamma_h(\tau \pm (\sigma_i - \sigma_j))}{\Gamma_h(\pm(\sigma_i -\sigma_j))} 
   \prod_{i=1}^{n} \prod_{a,b=1}^{f} \Gamma_h(\mu_a + \sigma_i;\nu_b -\sigma_i) \;.
\end{equation}
 The parameters $\omega_1$ and $\omega_2$ are associated to the squashing parameter $b$ by
$\omega_1=i b$, $\omega_2=i/b$. This can be used to simplify the formula above, fixing $\omega_1 \omega_2= -1$. 
The shorthand notations $\Gamma_h(x) \Gamma_h(y) = \Gamma_h(x;y)$ and
$\Gamma_h(x) \Gamma_h(-x) = \Gamma_h(\pm x)$ have been used in (\ref{Zgen}).
We will adopt the definition $\omega \equiv (\omega_1 + \omega_2)/2$ in the rest of the paper. 
We recall also a useful reflection equation, satisfied by the hyperbolic Gamma functions, that plays an important role in our analysis and it implies $\Gamma(\omega)=1$ as well,
\begin{equation}
\label{massive}
\Gamma_h (2 \omega - x) \Gamma_h(x) =1\,.
\end{equation}
Let us now explain the various terms appearing in the formula above;
The 
factor $n!$ corresponds to the dimension of the Weyl group.
The functions $\Gamma_h(x)$ appearing in the numerators
correspond to the one loop determinants of the
matter fields, while the ones appearing in the 
denominator are associated to the vector multiplet.
The arguments $x$ in $\Gamma_h(x)$  represent the linear combination of the
real scalars in the vector multiplets of the 
gauge symmetry, denoted as  $\sigma$, and of the weakly gauged global symmetries,
here denoted as $\mu,\nu$ and $\tau$.
They have to be taken in the weight of the representation for each
symmetry under which the fields transform. Note that these mass parameters for the weakly gauged global symmetries 
are generically complex, and the imaginary part represents the R-charge of 
each multiplet.

The partition function in equation (\ref{Zgen}) contains also the
contribution of a Fayet-Iliopoulos (FI) term  $ \lambda$,  which is computed as a classical contribution in localization.
In (\ref{Zgen}) we omit contributions of Chern-Simons (CS) terms to the partition function because we will not
consider them in our analysis.

Hyperbolic hypergeometric integrals as the one in formula (\ref{Zgen})  have been shown to satisfy various classes of integral identities.
It is remarkable to note that a  given integral can satisfy very different identities, depending on the constraints
satisfied by the complex parameters appearing in the argument of $\Gamma_h(x)$.
These constraints, defined as balancing conditions in the mathematical literature, translate on the physical playground into 
the presence of non-trivial superpotential interactions (often involving the presence of monopole operators).

It has been shown that 
a large quantity of such integral identities, most of them are listed in \cite{VanDeBult}, corresponds to the matching of the three
sphere partition function of 3d $\mathcal{N}=2$ models, and these have been used to corroborate or in other case to derive 3d $\mathcal{N}=2$ dualities.
However, there exist other identities discussed in \cite{VanDeBult} that have yet not been associated to any 3d duality.
In the following we will focus on some of these identities, discussing their connection with some dualities that appeared in the physics literatures recently.


\section{The Lagrangian $(A_1,A_{2n-1})$ theory and its reduction}
\label{sec:a1an}

In this section we review the 3d reduction of the $(A_1,A_{2n-1})$  AD theories derived 
in \cite{Benvenuti:2017lle,Benvenuti:2017kud} and show the analytic matching of its $S_b^3$ partition function
with the one of its mirror dual.
The starting point is the 4d construction of \cite{Maruyoshi:2016tqk,Maruyoshi:2016aim,Agarwal:2016pjo}.
One considers a 4d $SU(n)$ $\mathcal{N}=2$  theory with $2n$ flavors and couple them to 
a $2n \times 2n$ singlet $M$.
The superpotential of the theory is
\begin{equation}
\label{start4d}
W = \tr \, Q \Phi \tilde Q + \tr \, \tilde Q M Q\,.
\end{equation}
By assigning a vev to $\langle M \rangle$ that corresponds to the 
principal nilpotent orbit of the flavor symmetry
and by expanding around this vev
it has been shown that the theory flows to an IR fixed point if the contribution of accidental 
symmetries is included.

By following the prescription discussed in \cite{Kutasov:2003ux},
in order to modify a-maximization in presence of accidental symmetries.
In this case one should modify the Lagrangian
by adding some extra fields as discussed originally in \cite{Barnes:2004jj}.
The explicit use of such extra fields has appeared only very recently in \cite{Benvenuti:2017kud},
where the authors denoted them as flipping fields.
Supplementing this prescription with the 
chiral ring stability criterion, they obtained
 the 4d  $\mathcal{N}=1$  Lagrangian description of AD 
with superpotential
\begin{equation}
\label{WBG}
W =  \sum_{j=0}^{n-2} \alpha_i \tr \, q \phi^i \tilde q +  \sum_{j=2}^{n} \beta_i \tr \, \phi^i\,.
\end{equation}
This is the theory 
whose central charge coincides with the one obtained in the $(A_1,A_{2n-1})$ AD theory.

The 3d reduction of the theory mentioned above has been presented in \cite{Benvenuti:2017kud}. Note that unlike the discussion of 
\cite{Nii:2014jsa,Amariti:2014iza},
the Kaluza Klein (KK) monopole superpotential which is usually appearing when reducing 
4d dualities to 3d ones \cite{Aharony:2013dha}, is not generated here.
It is expected that the dimensional reduction of this theory is 
mirror dual to the reduction of the $(A_1,A_{2n-1})$ AD theories to 3d discussed in \cite{Nanopoulos:2010bv}.

Mirror symmetry relates theories (b) and (c) in Figure \ref{fig1} \cite{Intriligator:1996ex}.
It has been shown after performing F-maximization  that theory (a) is effectively equivalent to (b)\cite{Benvenuti:2017kud}. The exact R-charge for the adjoint field 
has been numerically found to be $r_\phi=0$, while the exact R-charge of the fundamentals
is $r_q = r_{\tilde q } = \frac{1}{2}$. The latter corresponds to the free field value, and it is a 
necessary result for the $\mathcal{N}=4$ hypermultiplets.

The net effect of having a zero $R$-charge for the adjoint is that the 
$SU(n)$ gauge symmetry \emph{abelianizes} into a $U(1)^{n-1}$ quiver.
On the partition function this can be understood because the one-loop determinant 
of the adjoint cancels out the one of the vector multiplet.
The abelian quiver is given in (b) with the following $\mathcal{N}=4$ superpotential 
\begin{equation}
W = \Phi_i (P_i \tilde P_i -P_{i+1} \tilde P_{i+1}),
\end{equation}
where $P_i$ and $\widetilde P_i$ form the $\mathcal{N}=4$ bifundamental hypermultiplets.
This quiver is mirror dual to the one in (c), corresponding to an $\mathcal{N}=4$ 
theory with superpotential 
\begin{equation}
\label{Wc}
W = \gamma_n \sum_{i=1}^{n} Q_i \widetilde{Q}_{n-i+1},
\end{equation}
where $Q_i$ and $\widetilde Q_{n-i+1}$ form the $\mathcal{N}=4$ hypermultiplets
and $ \gamma_n$ is a singlet.

Another crucial result of \cite{Benvenuti:2017kud} has been to obtain such a mirror dual description starting from the 
UV 3d model, obtained before integrating out the massive deformations and performing mirror symmetry at this stage. 
This gave rise to a 3d quiver theory with a series of nodes sequentially confining,
thanks to a new duality discovered in \cite{Benini:2017dud}.
The final quiver (c) has been obtained at the end of an intricate cascade of confinements.

As mentioned in the introduction, the direct connection between the models (a) and (c)
in Figure \ref{fig1} is still missing. In the following we will show that the equivalence of these two theories 
can be  obtained without any recursion to the ideas of \emph{abelianization} 
as well as sequential confinement of \cite{Benvenuti:2017kud}.
This would be a possible equivalence due to the analytical matching of their partition function for
generic values of the gauge rank $n$.
The matching is achieved by elaborating on an integral identity involving hyperbolic 
gamma functions mentioned in  \cite{VanDeBult}.

To prove the equivalence of the partitions functions in the models (a) and (c) $Z_{S_b^3}^{(a)} = Z_{S_b^3}^{(c)}$,  
we consider the {\bf Theorem  5.6.8} of \cite{VanDeBult} which states the following identity
\begin{eqnarray}
\label{vdb}
&&\frac{\Gamma_h(\tau)^n}{ n!}
\int \prod_{1 \leq j < k \leq n} \frac{\Gamma_h(\tau \pm (x_j -x_k))}{\Gamma_h(\pm(x_j-x_k))}
\prod_{i=1}^{n} \Gamma_h(\mu-x_i;\nu+x_i) 
e^{i \pi \lambda x_i }
dx_i
\nonumber \\
=&&
\prod_{j=0}^{n-1} \Gamma_h \bigg((j+1) \tau;j \tau+\mu+\nu;\omega-j \tau-\frac{\mu+\nu}{2} \pm \frac{\lambda}{2} \bigg)
e^{\frac{i \pi n \lambda}{2} (\mu- \nu)}.
\end{eqnarray}
%
%
%
Our first step would be to modify the $\Gamma_h$ functions appearing on the RHS of equation (\ref{vdb}). This modification will be
done by considering the following identities, calculated through  the reflection relation (\ref{massive})
\begin{eqnarray}
&&
\prod_{j=0}^{n-1} \Gamma_h ((j+1)\tau) = 
\frac{\Gamma_h(\tau)}{\prod_{j=1}^{n-1} \Gamma_h (2 \omega-(j+1)\tau)} 
=
\frac{\Gamma_h(\tau)}{\prod_{j=2}^{n} \Gamma_h (2 \omega- j \tau)}, 
\nonumber \\
&&
\prod_{j=0}^{n-1} \Gamma_h(j \tau+\mu+\nu)
=
\frac{\Gamma_h((n-1) \tau+\mu+\nu)}
{\prod_{j=0}^{n-2} \Gamma_h(2 \omega -j \tau-\mu-\nu)}.
\end{eqnarray}

The second step is integrating both sides of
identity (\ref{vdb}) over $\int d \eta$ where the paramters are related as $\eta=\frac{\lambda}{2} $.
On the field theory side 
this corresponds to turning on a vector multiplet (i.e. gauging) for the 
topological symmetry.
This gauging modifies the gauge group on the LHS
of equation (\ref{vdb}), converting the $U(n)$ factor into $SU(n)$ as done in
\cite{Park:2013wta,Aharony:2013dha,Aharony:2014uya}.
This can be seen by noting that the integral over $\eta$ on the LHS 
corresponds to $\delta \Big(\sum_{i=1}^{n} x_i \Big)$.
On the RHS of equation (\ref{vdb}) the integration over $\eta$
leaves a $U(1)$ gauge group with 
$n$ pairs of fundamentals and anti-fundamentals, corresponding to the
fields originally charged under the topological symmetry. We arrive at the following equality 
\begin{eqnarray}
\label{crucial}
&&\frac{\Gamma_h(\tau)^{n-1}}{ n!} 
\prod_{j=2}^{n} \Gamma_h (2 \omega- j \tau)
\prod_{j=0}^{n-2} \Gamma_h(2 \omega-j \tau-\mu-\nu)
\nonumber \\
\times &&
\int \prod_{1 \leq j < k \leq n} \frac{\Gamma_h(\tau \pm (x_j -x_k))}{\Gamma_h(\pm(x_j-x_k))}
\prod_{i=1}^{n} \Gamma_h(\mu-x_i;\nu+x_i) \delta\Big(\sum_{l=1}^{n} x_l\Big) dx_i
  \\
=&& 
\Gamma_h((n-1) \tau+\mu+\nu) \int d \eta
\prod_{j=1}^{n} \Gamma_h \bigg(\omega-(j-1) \tau-\frac{\mu+\nu}{2} \pm \eta\bigg)
e^{i \pi n \eta (\mu- \nu)}.
\nonumber 
\end{eqnarray}
Observe that the LHS is the 3d $\mathcal{N}=2$ 
$SU(n)$  theory with one flavor, one adjoint and the $\alpha_j$ and
$\beta_j$ singlets with superpotential 
(\ref{WBG}).
Indeed the one loop determinants of $\alpha_j$ and $\beta_j$, and the zero roots of the $SU(n)$ adjoint
appear in front of the integral, and the integrand has the flavors $q$ and $\tilde q$ with real masses
$\mu$ and $\nu$, respectively. The real mass for the adjoint has been identified with $\tau$.

The complex parameter $\mu$, $\nu$ and $\tau$ are unconstrained
\footnote{This translates on the field theory side into the 
absence of a superpotential for the KK monopoles.} and can be expressed as
\begin{eqnarray} 
\mu =   m_q + \omega \Delta_q, \quad
\nu =   m_{\tilde q} + \omega \Delta_{\tilde q}, \quad  
\tau =  m_{\phi} + \omega \Delta_{\phi}, 
\end{eqnarray}
where $m_i$ refers to the real mass of each field while $\Delta_i$ to its R-charge.

We will now prove that the RHS of equation (\ref{crucial}) corresponds to the theory obtained after sequential confinement and mirror symmetry. The dual theory is compatible with the superpotential (\ref{Wc}).
Our approach to prove this equality begins by studying the relation among the parameters $\mu$, $\nu$ and $\tau$.
Each superpotential term has $R$-charge $2$ and global charges $0$.
In the case at hand the charges of the field $\gamma_n$ can be read from the partition function
and we have
\begin{equation}
\mu_{\gamma_n} = (n-1) \tau+\mu+\nu
\end{equation}
while the $j$-th quark $Q_j$ and antiquarks $\tilde{Q}_j$ have charge
\begin{eqnarray}
\mu_{Q_j} &=& \omega-(j-1) \tau-\frac{\mu+\nu}{2} + \eta
\nonumber \\
\mu_{\widetilde Q_j} &=& \omega-(j-1) \tau-\frac{\mu+\nu}{2} - \eta
\nonumber
\end{eqnarray}
In this way each superpotential term is associated to the following  combination as expected
\begin{equation}
\mu_{\gamma_n}  + \mu_{Q_j}  + \mu_{\widetilde Q_{n+1-j}} = 2 \omega.
\end{equation}
The identity (\ref{crucial}) can be used to prove the results of \cite{Benvenuti:2017kud} with respect of
the enhancement of supersymmetry when $\Delta_\phi = 0 $ and $\Delta_{q} = \Delta_{\tilde q} = 1/2$.
In this case we can set $\tau=0$ and $\mu=\frac{\omega}{2}+b$, $\nu=\frac{\omega}{2}-b$.

Therefore, the RHS of equation (\ref{crucial}) becomes
\begin{equation}
\int d \eta\, 
e^{i \pi \,2 n  b \eta}\,
\Gamma_h \bigg( \frac{\omega}{2}\pm \eta\bigg)^n ,
\end{equation}
where the $FI$ terms is $ n b$.
On the other hand, the limit $\tau \rightarrow 0$ on the LHS is obtained using the following identity  \cite{Benvenuti:2017kud} 
\begin{equation}
\lim_{\tau \rightarrow 0 }
\Gamma_h(\tau) \Gamma_h (2 \omega- j \tau) = j,
\end{equation}
such that the final contribution of the first two terms in equation (\ref{crucial}) is $n!$, which
cancels the measure factor of $SU(n)$. In the integrand
the limit $\tau \rightarrow 0$ \emph{abelianizes} the gauge group, and one ends up with
the partition function of model $(b)$ as expected.
We conclude that the equivalence between the two sides of (\ref{crucial}) in this limit corresponds
to the equivalence between the partition function of the two $\mathcal{N}=4$ mirror
dual phases.

We can also use the identity  (\ref{crucial}) to study the $\mathcal{N}=2$ case
discussed in \cite{Benvenuti:2017kud}.
On the field theory side the difference consists of keeping the interaction 
\begin{equation}
\alpha_n q \phi^{n-1} \tilde q,
\end{equation}
in the  3d UV Lagrangian. By performing F-maximization in this case the
field $\alpha_n$ should not hit the unitary bound.
On the dual side the field $\gamma_n$ is massive, because of a mass term of the form 
$\gamma_n \alpha_n$.
Indeed the duality maps naturally the gauge invariant combination $q \phi^{n-1} \tilde q$
to the singlet $\gamma_n $. When $\alpha_n$ appears on one side corresponds to $\gamma_n$ 
disappearing on the other side, this fact is common in the Seiberg like dualities.

We can 
describe this mechanism on $Z_{S^3}$ by exploiting the relation 
(\ref{crucial}).
The field  $\alpha_n$ contributes to $Z_{S^3}$ with its one loop determinant.
It corresponds to multiply both sides of equation (\ref{crucial}) by 
$\Gamma_h(2\omega-(n-1) \tau-\mu-\nu)$.

Using the reflection relation  (\ref{massive}) on the RHS  of equation (\ref{crucial}) we obtain
\begin{equation}
\Gamma_h(2\omega-(n-1) \tau-\mu-\nu)  \Gamma_h((n-1) \tau+\mu+\nu) = 1. 
\end{equation}
The final result corresponds to the identity between the partition functions
of the expected $\mathcal{N}=2$ dual theories which discussed in \cite{Benvenuti:2017kud}.

\section{Comments on the $(A_1,D_4)$ model in 3d}
\label{sec:D4}

Another class of AD theories with an $\mathcal{N}=1$ Lagrangian description is 
denoted by $(A_1,D_{2n})$. 
In the case of even $n$ this theory has been constructed starting from the superpotential 
(\ref{start4d}), but with a vev $\langle M \rangle$ corresponding to a non-principal nilpotent orbit of the flavor symmetry group.
The final theory is $SU(n)$ SQCD with two flavors, an adjoint and a set of $\alpha_{j}$ and $\beta_{j}$ fields
interacting through the superpotential 
\begin{align}
\begin{split}
W = \sum^{n-2}_{j=0}\alpha_{j} \tr \, q \phi^{j} \tilde q + \tr \, p \phi \tilde p + \sum_{j=2}^{n}\beta_{j}\tr \,\phi^{j}\\
\end{split}
\end{align}
This model can be reduced to 3d, but in this case a monopole superpotential is generated \cite{Benvenuti:2017bpg}.
The reduction has been studied for the $(A_1,D_4)$ case and it has been shown that the theory 
is dual to an abelian gauge theory. In this case the \emph{abelianization} is not as simple as in the $(A_1,A_{2n-1})$ case,
essentially because the $R$-charge of the adjoint does not vanish.
In this case we have not been able to find any exact relation reproducing the matching of the original 
partition function with the mirror dual theory.

One possible way to have an analytical proof of the \emph{abelianization} at the level of the partition function 
consists of considering 
an $SU(2) \times U(1)$ quiver with one bifundamental flavor
connecting the two gauge groups and two flavors in the $SU(2)$ sector,
with superpotential
\begin{equation}
W = M \, \tr\, q_{12} q_{21} + \tr \, q_{21} q_{12} q_{2A} q_{A2} + s \, \tr \, q_{2B} q_{B2} + T_{U(1)}^{-}
\end{equation}
Here the indices $1$ and $2$ refer to the $U(1)$ and to the $SU(2)$ gauge groups, while 
$A$ and $B$ label the two flavors.
The field $T_-$ corresponds to the anti-monopole of the $U(1)$ gauge group.

This theory is dual to the model discussed above and can be shown as follows;
First we dualize the $U(1)$ node: it has two flavors and its dual is just given by
the meson $\Phi_{22} = q_{21} q_{12}$ interacting with a singlet $S_{+}$
(having the same charges of the monopole $T_{U(1)}^{-}$ of the electric theory).
This is one of the dualities derived in \cite{Collinucci:2017bwv,Benini:2017dud}.
The dual theory corresponds to $SU(2)$ with $2$ flavors and superpotential 
\begin{equation}
\label{Wduale}
W = \tr \, \Phi_{22} q_{2A} q_{A2} + s \, \tr \, q_{2B} q_{B2} + S_{+} \det{\Phi_{22}} + M \, \tr \, \Phi_{22} 
\end{equation}
The field $\Phi_{22}$ is a composite bifundamental field. It is made out of a singlet $\sim \phi \times  I_2 $ and an adjoint
$\Phi$.
By using the matrix identity $\det \Phi_{22} = -\frac{1}{2} \tr \, \Phi^2 + \phi^2 $ and by integrating out the massive 
fields $\phi$ and $M$ we can rewrite equation (\ref{Wduale}) as 
\begin{equation}
W = \tr \, \Phi_{22} q_{2A} q_{A2} + s \, \tr \, q_{2B} q_{B2} + S_{+}  \, \tr \, \Phi^2,
\end{equation}
corresponding to the superpotential of the $(A_1,D_4)$ theory reduced to 3d, 
in absence of the KK monopole superpotential. 
This  term can be turned on in both the phases without spoiling the duality just performed.

It should be possible directly prove of the \emph{abelianization} starting from the original $SU(2) \times U(1)$ quiver. In this case the absence of
adjoint matter simplifies the problem and it allows to use a larger web of 3d $\mathcal{N}=2$
dualities. We hope to come back to this issue in the  future.
\section{Further applications}
\label{sec:further}
In this section we discuss some further examples of integral identities
involving hyperbolic hypergeometric integrals.
The identities that we will discuss are listed in \cite{VanDeBult} and they represent
symplectic gauge groups with matter fields in the 
fundamental and in the antisymmetric representations.
Here we restrict our analysis to the case of $SP(2)=SU(2)$ gauge group, 
where the antisymmetric field disappears.
We show that in the case with six fundamentals the integral identity reduces
to a modification of Aharony duality studied in \cite{Benini:2017dud}, with a
quadratic monopole superpotential.
 We recover the identity for a  limiting case of Aharony duality for the case of four fundamentals.%
\subsection{An exact relation for a higher power monopole duality}
\label{subsec:multi}

Let us discuss one of the new dualities found in \cite{Benini:2017dud},
involving power monopole superpotentials. 
We show that, when the gauge group is $U(2)$, the matching of the
partition functions of the dual theories
can be derived from  {\bf Formula 5.3.7} of \cite{VanDeBult}:
\begin{eqnarray}
\label{vdb1}
&&
\frac{\Gamma_h(\tau)^n}{ 2^n n! }
\int \prod_{i=1}^{n}
 dx_i
 \prod_{1 \leq i < l \leq n} \frac{\Gamma_h(\tau \pm x_i \pm x_l)}{\Gamma_h(\pm x_i \pm x_l)}
\prod_{i=1}^{n}
\frac{\prod_{a=1}^{6} \Gamma_h(\mu_a \pm x_i)}{\Gamma_h( \pm 2 x_i)}
\nonumber \\
=&&
\prod_{j=0}^{n-1} \Gamma_h((j+1)\tau) 
\prod_{a<b} \Gamma_h(j \tau + \mu_a + \mu_b).
\end{eqnarray}
This identity holds provided that the complex  parameters $\tau$ and $\mu_a$
satisfy the balancy condition
\begin{equation}
\label{bch}
2(n-1) \tau + \sum_{a=1}^{6} \mu_a = 2 \omega.
\end{equation}
The LHS of equation (\ref{vdb1}) corresponds to an $SP(2n)$ theory with an antisymmetric and six fundamentals. Furthermore, it is an interesting observation that (\ref{vdb1}) can be obtained in a physical way. This follows from 
the circle reduction of a limiting case of "rank changing" dualities for symplectic gauge theories with 
eight fundamentals and an antisymmetric, recently discovered in \cite{Razamat:2017hda}.
This observation deserves further studies and it may be shed some light on the four dimensional origin of
the dualities with monopole superpotentials having higher powers discovered in \cite{Benini:2017dud}.

Here we restrict to the case $n=1$. In this case the measure factor
simplifies leaving an $SU(2)$ theory. 
We further choose the mass parameters as
\begin{equation}
\mu_i = m_i + m_B +m_A, 
\quad
\mu_{i+3} = n_i - m_B + m_A 
\quad
\text{for } i=1,2,3
\end{equation}
with the further constraints $\sum_i {m_i}=\sum_i {n_i} = 0$.
This fixes $m_A = \frac{\omega}{3}$, that will be crucial in the following.
So far we are just reassembling the real masses, with a parameterization 
compatible with a global $SU(3)_L \times SU(3)_R \times U(1)_B$
symmetry.
We can modify the $SU(2)$ gauge symmetry to $U(2)$ by gauging the global $U(1)_B$ symmetry .
This gauging corresponds to integrate both sides of equation (\ref{vdb1}) over $\int  d m_B$.
On the LHS we also re-define the integration variables as $m_B+x = \sigma_1$ and $m_B-x = \sigma_2$,
while we denote $m_B = \sigma$ on the RHS.
Thus we arrive at the following identity 
\begin{eqnarray}
\label{newrel}
&&
\int 
\prod_{i=1}^{2} 
d \sigma_i 
  \frac{ \prod_{a=1}^{3} 
\Gamma_h(m_a+ m_A + \sigma_i;n_a+m_A  - \sigma_i)}{
\Gamma_h (\pm(\sigma_1 - \sigma_2))}
\\
&& =
\prod_{a,b} \Gamma_h(m_a + n_b + 2 m_A)
\int d \sigma
 \prod_{a=1}^{3}
\Gamma_h( 2m_A-m_a  +\sigma; 2m_A-n_a  -\sigma).
\nonumber 
\end{eqnarray}

This relation looks similar to the one expected for an Aharony  duality 
between a $U(2)$ and a $U(1)$ theory with three flavors, but without 
electric monopoles acting as singlets and constraining the chiral ring of the dual phase.
Moreover, differently from the ordinary Aharony duality, here 
the dual quarks have real mass $2 m_A$ instead of $-m_A$.

In order to study the field theory properties of the duality underlining the identity
(\ref{newrel}),  we make use of the constraint $m_A = \omega/3$. This signals the fact that the 
real part of $m_A$ vanishes, while its imaginary part, corresponding to the trial R-charge, is fixed. This is 
also the exact R-charge, because of the non-abelian nature of the other global symmetries.
 In this case, the duality map implies that the electric quarks have $R$-charge $\Delta_Q$ while the 
magnetic quarks have $R$-charge $\Delta_q=2 \Delta_Q$. It is different from the 
 expected one in the ordinary Aharony duality, $1-\Delta_Q$.
However, the balancing condition gives $\Delta_Q = \frac{1}{3}$, compatible with 
$\Delta_Q = 1-\Delta_q$.

We conclude the analysis by checking that, when the real masses and the R-charges are constrained by condition (\ref{bch}), the relation (\ref{newrel})
corresponds to the new duality conjectured in  \cite{Benini:2017dud}, involving monopole
superpotentials with quadratic powers. 
This duality has been formulated for 3d $\mathcal{N}=2$ $U(n)$ SQCD with $f$ flavors and 
superpotential 
\begin{equation}
W = T_+^2 + T_-^2.
\end{equation}
The dual theory corresponds to 3d $\mathcal{N}=2$ $U(f-n)$ SQCD with $f$ dual flavors and 
\begin{equation}
W = M q \tilde q +t_+^2 + t_-^2.
\end{equation}
The presence of a monopole superpotential imposes that 
\begin{equation}
\label{monch}
\Delta_{T_{\pm}} = f (1-\Delta_Q) - n + 1 = 1,
\quad
\Delta_{t_{\pm}} = f (1-\Delta_q) - \widetilde{n} + 1 = 1.
\end{equation}

In our analysis we have studied the case with $n=2$ and $f=3$. In this case the duality of  
\cite{Benini:2017dud} implies $\widetilde n=1$. Plugging these values in (\ref{monch}) we 
find
\begin{equation}
\Delta_Q = \frac{1}{3},\quad  \quad \Delta_{q} = \frac{2}{3}
\end{equation}
that corresponds to the values obtained above from the analysis of the partition functions.
\subsection{A limiting case of Aharony duality}
\label{subsec:limaha}
As the last example of the  hyperbolic identities with application to the 3d $\mathcal{N}=2$ dualities we 
discuss the identity corresponds to  {\bf Theorem 5.6.6}
of \cite{VanDeBult} as follow
\begin{eqnarray}
\label{vdb2}
&&
\frac{\Gamma_h(\tau)^n}{ 2^n n!  }
\int 
\prod_{i=1}^{n}
 dx_i
\prod_{1 \leq i < l \leq n} \frac{\Gamma_h(\tau \pm x_i \pm x_l)}{\Gamma_h(\pm x_i \pm x_l)}
\prod_{i=1}^{n}
\frac{\prod_{a=1}^{4} \Gamma_h(\mu_a \pm x_i)}{\Gamma_h( \pm 2 x_i)}
\nonumber \\
=&&
\prod_{j=0}^{n-1} \frac{\Gamma_h((j+1)\tau)}{\Gamma_h((2n-2-j)\tau+\sum_r \mu_r)} 
\prod_{a<b} \Gamma_h(j \tau + \mu_a + \mu_b).
\end{eqnarray} 
Same as previous example, we fix $n=1$ and then we observe that this identity can be obtained as a limiting case of equation (\ref{vdb1}).
This corresponds to a real mass flow on the field theory side.

In addition, we parameterize the real masses $\mu$ as 
\begin{equation}
\mu_a = m_a + m_B +m_A, 
\quad
\mu_{a+2} = n_a - m_B + m_A 
\quad
\text{for } a=1,2
\end{equation}
with the further constraints $\sum_i {m_i}=\sum_i {n_i} = 0$.
So far we are just reassembling the real masses, with a parameterization 
compatible with a global $SU(2)_L \times SU(2)_R \times U(1)_B \times U(1)_A$
symmetry.
The $SU(2)$ gauge symmetry becomes $U(2)$ provided that we gauge the $U(1)_B$ symmetry
and add the FI term.
This is done by integrating both sides of the identity by
\begin{equation}
\int d m_B e^{4 i \pi \lambda m_B},
\end{equation}
where the normalization on the FI is arbitrarily chosen for the future purposes.

Redefine the integration variables as $m_B+x = \sigma_1$ and $m_B-x = \sigma_2$ we get
\begin{eqnarray}
\label{Zint}
\int  \prod_{i=1}^{2} d \sigma_i  e^{2 \pi  i \lambda x_i  }
\!\!&&\!\!
\prod_{a=1}^{2} 
\Gamma_h(m_a+ m_A + \sigma_i) \Gamma_h(n_a+m_A  - \sigma_i)
\Gamma_h^{-1} (\pm(\sigma_1 - \sigma_2))
\nonumber  \\
&& =
\frac{\prod_{a,b} \Gamma_h(m_a + n_b + 2 m_A)}{\Gamma_h(4m_A)}
\int d \sigma e^{2  i \pi \lambda \sigma}
\Gamma_h( 2m_A  \pm \sigma).
\end{eqnarray}
The integral on the RHS corresponds to the partition function of $U(1)$ with
two flavors, dual to the XYZ model \cite{Jafferis:2010un}. In this case we can use the following identity 
\begin{equation}
\label{XYZ}
\int d \sigma  e^{2 i \pi \lambda \sigma}
\Gamma_h( 2m_A  \pm \sigma) 
=
\Gamma_h(4m_A) \Gamma_h\big(\pm \frac{\lambda}{2}-2 m_A \big),
\end{equation}
%
and substituting it into (\ref{Zint}) and get the following equality
\begin{eqnarray}
\label{final}
&&
\int 
 \frac{\prod_{i=1}^{2} d \sigma_i e^{2 \pi  i \lambda x_i  }\prod_{a=1}^{2} 
\Gamma_h(m_a+ m_A + \sigma_i) \Gamma_h(n_a+m_A  - \sigma_i)}{
\Gamma_h (\pm(\sigma_1 - \sigma_2))}
\\
&& =
\prod_{a,b} \Gamma_h(m_a + n_b + 2 m_A)
\Gamma_h\big(\pm \frac{\lambda}{2}-2 m_A \big).
\nonumber 
\end{eqnarray}
This equality corresponds to the limiting case of Aharony duality for the $U(2)$ model with two flavors \cite{Aharony:1997gp}.
Indeed in this case the expected dual has the following superpotential
\begin{equation}
W = v_+ v_- \det M = v_+ v_- (M_{11} M_{22} - M_{12} M_{21}).
\end{equation}
We can observe that the constraints imposed from this superpotential 
are exactly encoded in (\ref{final}).
The FI term $\lambda$ corresponds to the real mass for the monopoles $v_{\pm}$ and
the $SU(2)_L \times SU(2)_R \times U(1)_A$ global charges reproduce the 
monopole and the mesons contributions.
\section{Conclusions}
\label{sec:conc}
In this paper we studied 3d $\mathcal{N}=2$ theories arising from the 
reduction of 4d $\mathcal{N}=1$ Lagrangian theories, conjectured to 
enhance to $(A_1,A_{2n-1})$ AD theories.
We provided a check of the IR duality relating the model reduced to 3d
and its mirror dual, matching the three sphere partition functions. 
This corroborates the duality claimed among these models in
\cite{Benvenuti:2017lle,Benvenuti:2017kud}.
This check has been possible thanks to an integral identity, listed in \cite{VanDeBult},
in terms of hyperbolic hypergeometric integrals.
Meanwhile, we did not find an analogous relation for the $(A_1,D_4)$
case. 
We studied also other identities, associated to $SP(2n)$ gauge theories
with fundamentals and an antisymmetric, showing that in the $n=1$ case
they reduce to known 3d $\mathcal{N}=2$  dualities.

We believe that two main aspects of our analysis require a deeper analysis and 
may lead to interesting results.
The first aspects regards the $(A_1,D_4)$ case. In section \ref{sec:D4} we have obtained 
a dual description of the Lagrangian reduction of the $\mathcal{N}=1$ theory that enhances to 
AD. This 3d duality can be helpful in proving the abelianization and it 
deserves further investigations.
Another interesting connection that emerged in the analysis regards the relation between the identity 
(\ref{vdb1}) and the dualities proposed in \cite{Razamat:2017hda}, based on the results of \cite{Rains}.  
We have seen how the relation can be interpreted, in a simplified case in absence of antisymmetric matter,
in terms of the dualities of \cite{Benini:2017dud} involving quadratic powers in the monopole superpotential.
It may be interesting to develop a more general analysis for the 3d dualities obtained from the $S^1$  reduction of 
\cite{Razamat:2017hda} and further real mass and Higgs flow. We hope to report on progress in this direction
in the next future.

\section*{Acknowledgements}
We are grateful to Sergio Benvenuti and Domenico Orlando for interesting and stimulating discussions.
The work of N.A. and Y.S. is supported by the 
Swiss National Science Foundation (\textsc{snf}) under grant number \textsc{pp}00\textsc{p}2\_157571/1.
The work of A.A. is supported by INFN.
\bibliographystyle{JHEP}
\bibliography{references}

\end{document}